\documentclass[twocolumn]{aastex63}
\usepackage{epsfig}
\usepackage{amsmath, amsthm, amssymb}
\usepackage{microtype}
\usepackage{float}
\usepackage{verbatim}
\usepackage{wrapfig}
\usepackage{graphicx}
\usepackage[caption = false]{subfig}
\usepackage{multirow}
\usepackage{hhline}
\usepackage{booktabs}
\usepackage{array}
\usepackage{bm}
\usepackage[normalem]{ulem}

\accepted{\today}

\shorttitle{A new candidate transitional MSP}
\shortauthors{Miller et al.}

\begin{document}

\title{A New Candidate Transitional Millisecond Pulsar in the Sub-luminous Disk State: 4FGL J0407.7--5702}

\correspondingauthor{Jessie M. Miller}
\email{mill2890@msu.edu}

\author{Jessie M. Miller}
\affiliation{Center for Data Intensive and Time Domain Astronomy, Department of Physics and Astronomy,\\ Michigan State University, East Lansing, MI 48824, USA}
\author{Samuel J. Swihart}
\affiliation{Center for Data Intensive and Time Domain Astronomy, Department of Physics and Astronomy,\\ Michigan State University, East Lansing, MI 48824, USA}
\author{Jay Strader}
\affiliation{Center for Data Intensive and Time Domain Astronomy, Department of Physics and Astronomy,\\ Michigan State University, East Lansing, MI 48824, USA}
\author{Ryan Urquhart}
\affiliation{Center for Data Intensive and Time Domain Astronomy, Department of Physics and Astronomy,\\ Michigan State University, East Lansing, MI 48824, USA}
\author{Elias Aydi}
\affiliation{Center for Data Intensive and Time Domain Astronomy, Department of Physics and Astronomy,\\ Michigan State University, East Lansing, MI 48824, USA}
\author{Laura Chomiuk}
\affiliation{Center for Data Intensive and Time Domain Astronomy, Department of Physics and Astronomy,\\ Michigan State University, East Lansing, MI 48824, USA}
\author{Kristen C. Dage}
\affiliation{Center for Data Intensive and Time Domain Astronomy, Department of Physics and Astronomy,\\ Michigan State University, East Lansing, MI 48824, USA}
\author{Adam Kawash}
\affiliation{Center for Data Intensive and Time Domain Astronomy, Department of Physics and Astronomy,\\ Michigan State University, East Lansing, MI 48824, USA}
\author{Laura Shishkovsky}
\affiliation{Center for Data Intensive and Time Domain Astronomy, Department of Physics and Astronomy,\\ Michigan State University, East Lansing, MI 48824, USA}
\author{Kirill V. Sokolovsky}
\affiliation{Center for Data Intensive and Time Domain Astronomy, Department of Physics and Astronomy,\\ Michigan State University, East Lansing, MI 48824, USA}
\affiliation{Sternberg Astronomical Institute, Moscow State University, Universitetskii pr. 13, 119992 Moscow, Russia}
\affiliation{Astro Space Center of Lebedev Physical Institute, Profsoyuznaya St. 84/32, 117997 Moscow, Russia}


\begin{abstract}

We report the discovery of a variable optical and X-ray source within the error ellipse of the previously unassociated \textit{Fermi} Large Area Telescope $\gamma$-ray source 4FGL J0407.7--5702.
A 22 ksec observation from \emph{XMM-Newton}/EPIC shows an X-ray light curve with rapid variability and flaring. The X-ray spectrum is well-fit by a hard power law with $\Gamma = 1.7$. Optical photometry taken over several epochs is dominated by aperiodic variations of moderate amplitude. Optical spectroscopy with SOAR and Gemini reveals a blue continuum with broad and double-peaked H and He emission, as expected for an accretion disk around a compact binary. Overall, the optical, X-ray, and $\gamma$-ray properties of 4FGL J0407.7--5702 are consistent with a classification as a transitional millisecond pulsar in the sub-luminous disk state. We also present evidence that this source is more distant than other confirmed or candidate transitional millisecond pulsar binaries, and that the ratio of X-ray to $\gamma$-ray flux is a promising tool to help identify such binaries, indicating that a more complete census for these rare systems is becoming possible.\\

\end{abstract}

\vspace{2mm}
\section{Introduction}

Transitional millisecond pulsars (tMSPs) form a unique subclass of stellar compact binary system containing a neutron star and a low-mass, non-degenerate companion. These binaries are identified by their observed transitions between a low-mass X-ray binary-like state and a rotation-powered radio millisecond pulsar state on timescales of weeks to years. To date, only three systems have been observed to undergo such transitions: PSR J1023+0038 \citep{Archibald09} and XSS J12270--4859 \citep{Bassa14,Roy15} in the Galactic field, and IGR J18245--2452 \citep{Papitto13} in the globular cluster M28. 

In addition to these class-defining transitions, tMSPs show other distinctive characteristics, especially in the disk state. Their typical 1--10 keV X-ray luminosities are $L_X\sim10^{33-34}$ erg s$^{-1}$ \citep{deMartino13,Papitto13,Patruno14}. This is much lower than the typical values of $L_X \gtrsim 10^{36}$ erg s$^{-1}$ observed for persistent or outbursting transient neutron star low-mass X-ray binaries \citep{Paradijs98}. Hence this $L_X\sim10^{33-34}$ erg s$^{-1}$ state is also called the sub-luminous disk state. 
A single tMSP (IGR J18245--2452; \citealt{Papitto13}) has been observed at the outburst 
$L_X$ of a few $\times \,  10^{36}$ erg s$^{-1}$ traditionally seen for accreting millisecond X-ray pulsars \citep{Patruno12}. While it is uncertain what fraction of tMSPs show such outbursts, this case shows that it is a possible alternative route to tMSP discovery.

\begin{figure*}[ht!]
\begin{center}
	\includegraphics[width=0.45\linewidth]{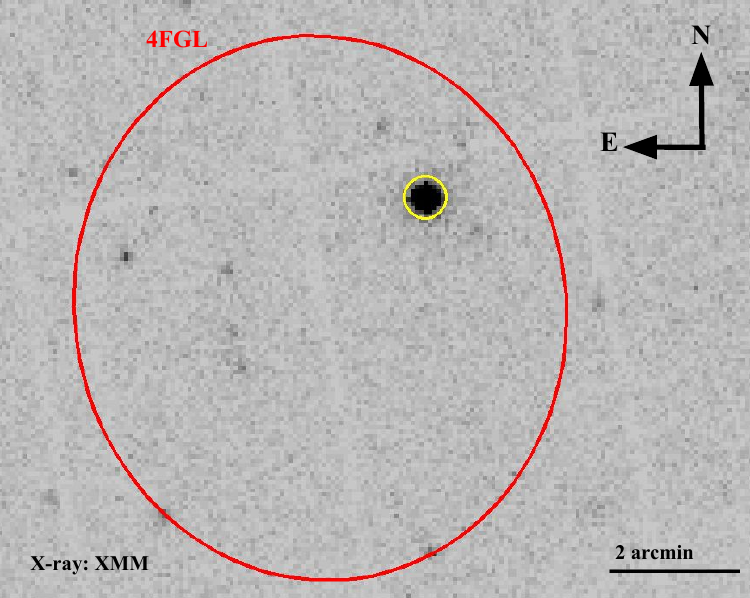}
	\hspace{5mm}
	\includegraphics[width=0.45\linewidth]{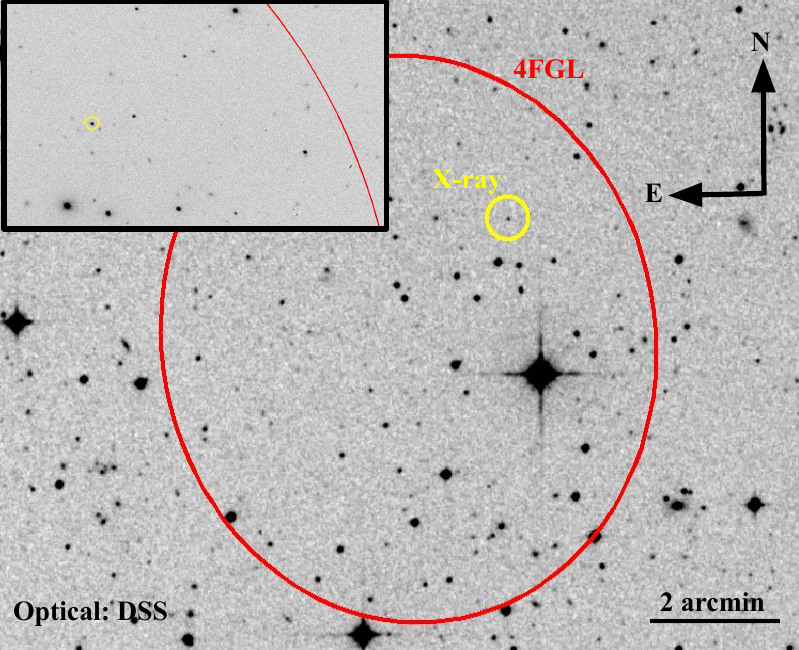}
    \caption{Left: \textit{\textit{XMM-Newton}}/EPIC  X-ray image of the field of 4FGL J0407.7--5702. The 95\% 4FGL error ellipse is in red and the candidate X-ray/optical counterpart is circled in yellow. Right: Red Digitized Sky Survey image of the field with the X-ray source marked in yellow. The inset is a zoomed-in unfiltered optical image taken with the SOAR telescope, with the source again marked with a yellow circle.}
\end{center}
\label{fig:finder_fig}
\end{figure*}

The X-ray light curves show variability on a range of amplitudes and timescales. In the three confirmed tMSPs this variability exhibits remarkable ``mode switching": abrupt changes between distinct low and high modes \citep{deMartino13,Linares14,Bogdanov15} that differ by a factor of $\sim 5-10$ in X-ray luminosity, with occasional more luminous flares. During the high mode there is evidence of X-ray and optical pulsations \citep{Archibald10,Archibald15,Papitto2015b,Ambrosino17}. In the low mode X-ray pulsations are not seen, and instead radio continuum emission is observed, likely from synchrotron-emitting bubbles produced close to the neutron star \citep{Bogdanov18}.

A self-consistent model of these observations has not yet been firmly established. Indeed it is unclear whether (i) the disk state is accretion powered, with some accreted material reaching the surface of the neutron star and the rest possibly ejected in a propeller \citep{Bogdanov15,Papitto2015a} or maintained in a ``trapped disk" \citep{Dangelo12}, or (ii) whether the disk state is instead rotation powered, with the X-ray emission (including the X-ray pulsations) due to shocks from the pulsar wind occurring just outside the light cylinder \citep{Ambrosino17,Papitto19,Veledina19} or at a larger radius \citep{Takata14}. Notably, timing observations of PSR J1023+0038 show that the pulsar spin-down rate increases by $\sim 27\%$ in the sub-luminous disk state compared to the radio pulsar state \citep{Jaodand16}. This suggests that the pulsar wind is enhanced during the disk state, rather than being suppressed, as might be expected if accretion is occurring. The detection of radio pulsations in the disk state could help settle the question, but these have not been detected in the sub-luminous disk state of any of the confirmed tMSPs despite extensive searches \citep{Hill11,Papitto13,Stappers14,Jaodand16}.

In another departure from typical low-mass X-ray binary phenomenology, the two field tMSPs show enhanced GeV $\gamma$-ray emission in the sub-luminous disk state: they are brighter by a factor of a few compared to the millisecond pulsar state \citep{Stappers14,Johnson15}. Besides the known tMSPs (and the candidate tMSPs selected via $\gamma$-ray emission), no other low-mass X-ray binary at $L_X \lesssim 10^{35}$ erg s$^{-1}$ has shown persistent $\gamma$-ray emission, including a number of relatively nearby accreting millisecond X-ray pulsars \citep{Torres20}.

Regardless of whether the disk state in tMSPs is actually powered by accretion, there is strong evidence from the optical that a disk is indeed present. Optical spectra in this state are dominated by a warm continuum and the double-peaked H and He emission lines characteristic of an accretion disk around a compact object \citep{Bond02,deMartino14}. In the MSP state, these optical signatures of a disk disappear \citep{Thorstensen05,deMartino15}.

Unlike the apparently unique sub-luminous disk state, tMSPs in the pulsar state have multiwavelength properties comparable to the other compact binaries with non-degenerate companions. These systems are typically called ``redbacks" when the companion mass is $\gtrsim 0.1 M_{\odot}$ \citep{Roberts13}. The X-ray emission in redbacks is ususally dominated by a hard intrabinary shock modulated on the orbital period \citep{Linares_solo14,Roberts14}. Typical pulsar searches strongly select against eclipsing pulsars such as redbacks, and targeted follow-up of new $\gamma$-ray sources discovered with the \emph{Fermi} Large Area Telescope has been the most important tool in substantially increasing the known sample of field redbacks \citep{Ray12,Roberts13}.

Redbacks are not mere curiosities: as a population they have among the highest neutron star masses known for any subclass of millisecond pulsars \citep{Strader19}, and some are among the fastest spinning pulsars known \citep{Patruno17}. Since all tMSPs are redbacks it is natural to wonder whether the converse holds, but despite intensive searches (e.g., \citealt{Torres17}) no other field redbacks have been observed to transition to or from the pulsar state.

Instead, new candidate tMSPs have been identified using the distinguishing characteristics of the class in the sub-luminous disk state, including $\gamma$-ray emission, variable X-ray emission with $L_X \sim 10^{33}-10^{34}$ erg s$^{-1}$, and evidence for a disk. The three convincing candidates found this way are: 3FGL J1544.6--1125 \citep{BH15}, 3FGL J0427.9--6704 \citep{Strader16}, and CXOU J110926.4--650224 \citep{CotiZelati19}. This tMSP discovery route, while very useful, is certainly incomplete: some other sources have X-ray luminosities and variability properties somewhat similar to tMSPs in the sub-luminous disk state, but perhaps have not been detected as $\gamma$-ray sources due to their distances or confused sky locations (e.g., \citealt{Degenaar14,Heinke15}). It is plausible that some of these sources are indeed tMSPs and could be identified as such with future multi-wavelength observations.

Here we present the discovery and characterization of a compact binary within the error ellipse of the \emph{Fermi}-LAT $\gamma$-ray source 4FGL J0407.7--5702. We show this source has X-ray and optical properties similar to the known tMSPs in the sub-luminous disk state
and hence is a strong tMSP candidate.

\section{Observations}

\subsection{The $\gamma$-ray Source \& Optical Discovery}
\label{sec:gamma}

The candidate X-ray/optical counterpart was discovered as part of our ongoing program to search for new compact binaries among previously unassociated \emph{Fermi}-LAT $\gamma$-ray sources. We are focused in particular on possible counterparts that are both optical variables and X-ray sources, since this preferentially selects for compact binaries over unrelated contaminants.

The focus of this paper is on an X-ray/optical source at the edge of the 68\% error ellipse---and hence well within the 95\% error ellipse---of the LAT 4FGL-DR2 (10-year) source 4FGL J0407.7--5702 \citep{4FGLDR2}. The 95\% error ellipse is not too far from circular, with a mean radius $\sim 4.1$\arcmin. The LAT source, while faint (0.1--100 GeV flux of 1.6$\pm$0.3 $\times 10^{-12}$ erg s$^{-1}$ cm$^{-2}$), is detected at $5.7\sigma$, with a power law photon index of $\Gamma = 2.54\pm0.17$.
There is no significant evidence for variability either in the formal variability index or in an examination of the light curve in 1-yr bins \citep{4FGLDR2}.

Using archival \textit{Swift} X-ray Telescope (XRT) data \citep{Stroh13} taken from the \emph{Swift}/XRT website \citep{Evans09}, we identified a single prominent X-ray source within the \emph{Fermi}-LAT error ellipse, with a J2000 (R.A., Dec.) of (04:07:31.78, --57:00:25.2) and a 90\% uncertainty of 4.5\arcsec. There is a single catalogued optical source that matches this X-ray source. The \emph{Gaia} DR2 ICRS position of this source in (R.A., Dec.) is (04:07:31.7195, --57:00:25.295), which we take as the best known position. This match is $< 1$\arcsec, so well within the uncertainty of the X-ray position, and the follow-up data discussed below prove the X-ray and optical sources are associated with each other. Furthermore, the \emph{Gaia} DR2 photometry ($G = 20.176\pm0.011$ mag), and presence of the optical source in a catalog of candidate variables identified by the Dark Energy Survey \citep{Stringer19}, both suggested it might be variable, motivating follow-up.

\subsection{Optical Spectroscopy}
\label{sec:optobs}

We obtained optical spectroscopy with the Goodman Spectrograph \citep{Clemens04} on the SOAR telescope on parts of six different nights from 2019 Nov 6 to 2020 Jan 17. In all cases we used a 400 l mm$^{-1}$ grating with a 0.95\arcsec\ slit, giving a resolution of about 5.6 \AA\ (full-width at half-maximum; FWHM). Some of the spectra were obtained with a wavelength range of $\sim 3820$--7850 \AA, while others used a central wavelength with coverage about 1000 \AA\ redder. Each spectrum had an exposure time of 1500 sec. The spectra were all reduced and optimally extracted in the normal manner.

We also obtained several spectra with Gemini/GMOS-S (Program ID: GS-2019B-FT-111) on the nights of 2019 Dec 30 and 31. On each night three 1200 sec exposures were taken. The R400 grating and a 1.0\arcsec\ slit together yielded a FWHM resolution of about 7.2 \AA\ over the wavelength range $\sim 4500$--9150 \AA. These data were reduced using the Gemini IRAF package \citep{2016ascl.soft08006G}.

\subsection{SOAR Photometry}
\label{sec:photometry}
In an effort to detect periodic optical variability associated with the companion, we observed the source with SOAR/Goodman in imaging mode on 2019 Dec 16 and again on 2020 Jan 12. On 2019 Dec 16, which had seeing around 1\arcsec, we took a series of 180 s exposures, alternating between the SDSS $g'$ and $i'$ filters, while on the second, brighter night, with median seeing of 0.8\arcsec, we only observed in $i'$, with a frame time of 120 s. We performed differential aperture photometry with respect to a set of nearby non-varying stars, and calibrated to magnitudes from the Dark Energy Survey \citep{DES18}.

\subsection{X-ray Observations}
\label{sec:xrayobs}
We observed 4FGL J0407.7--5702 on 2020 March 6 with the European Photon Imaging Camera (EPIC) on board the \textit{XMM-Newton} space telescope. A total live time of $\sim 22$~ksec was achieved. Data were reduced using the Science Analysis Software ({\sc sas}) data reduction package, version 18.0.0. We used a circular source extraction region of radius 30$^{\prime\prime}$ centered on J0407.7--5702 and a local background extraction region with an area three times larger. Exposure time intervals of high particle backgrounds were excluded. Standard flagging criteria \verb| FLAG=0|, plus \verb|#XMMEA_EP| and \verb|#XMMEA_EM| (for pn and MOS detectors, respectively), were applied. Additionally, we selected patterns 0--4 for the pn data and 0--12 for the MOS data. Individual background-subtracted spectra were extracted for pn, MOS1 and MOS2 using standard tasks in \textit{xmmselect}. A single combined EPIC spectrum was created using \textit{epicspeccombine} and grouped to at least 20 counts per bin so that Gaussian statistics could be used. For our timing analysis, we used the {\sc sas} tasks \textit{evselect} and \textit{epiclccorr} to produce a background-subtracted light curve.

The X-ray source discussed in this paper is the closest source to the center of the \emph{Fermi}-LAT error ellipse, and has a higher X-ray flux than any other source in the error ellipse by a factor of $\sim 20$.

\section{Results \& Analysis}

\subsection{X-ray Spectrum and Mean X-ray Properties}
\label{sec:Xrayspec}

\begin{figure}[t!]
	\includegraphics[width=\linewidth]{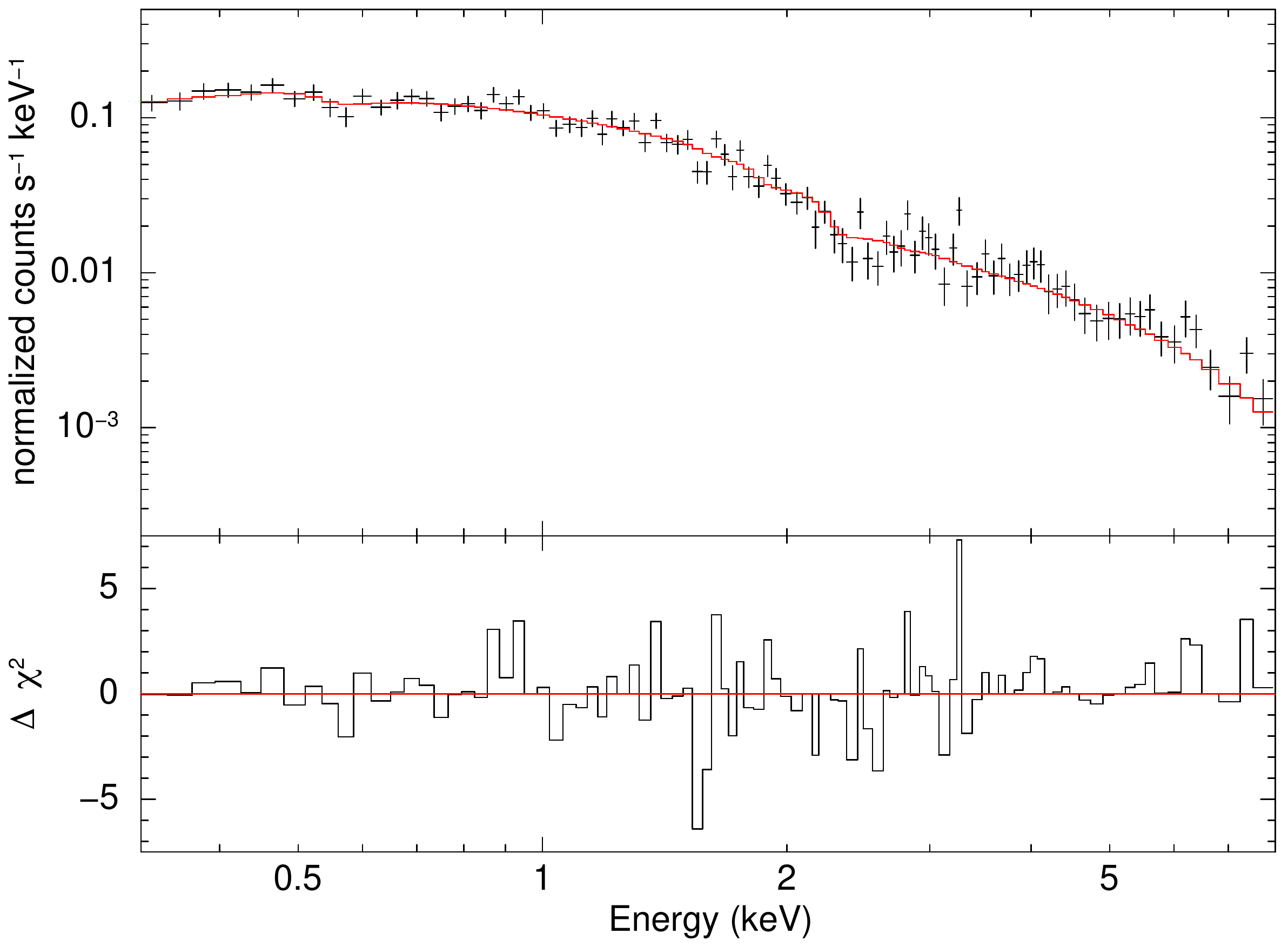}
    \caption{\emph{XMM-Newton}/EPIC spectrum of 4FGL J0407.7--5702, which is well-fit by an absorbed power law with $\Gamma=1.74\pm0.04$.}
\label{fig:xray_spectra}
\end{figure}

We began by fitting the \textit{XMM-Newton} EPIC spectrum with an absorbed power-law, as shown in Figure~\ref{fig:xray_spectra}. At the Galactic latitude of 
the source ($b = -44^{\circ}$) the expected foreground extinction is very low ($E(B-V) = 0.013$; \citealt{Schlafly11}), with a correspondingly low line-of-sight column density of  $N_H=1.08\times10^{20}$ cm$^{-2}$ \citep{HI4PI16}. We find no evidence of additional intrinsic absorption and so fix the $N_H$ to this foreground value.
This satisfactory model ($\chi^2/$dof$=105.8/96$, $p=0.23$) results in a best-fit photon index of $\Gamma=1.74\pm0.04$, with an unabsorbed 0.5--10 (1--10) keV flux of $4.16\pm0.17$ ($3.47\pm0.17$) 
$\times10^{-13}$ erg s$^{-1}$ cm$^{-2}$. At a reference distance of 7 kpc (see Section~\ref{sec:distance}), this flux corresponds to a 0.5--10 keV X-ray luminosity of $2.4 \, (d/7 \, {\rm kpc})^2 \times 10^{33}$ erg s$^{-1}$.

We also experimented with adding a blackbody component to the model. However, we found only a marginal improvement ($\chi^2$/dof=101.6/94) in the fit, with an $F$-test probability of $p=0.15$.
As expected, in this fit, the added blackbody component (\textit{k}T=0.17$\pm$0.05 keV) results in a slightly harder power law component ($\Gamma=1.63\pm0.10$), but the blackbody contributes only $\sim 5$\% to the unabsorbed 0.5--10 keV flux. For this two-component model, the unabsorbed 0.5--10 (1--10) keV flux is $4.3\pm0.2$ ($3.6\pm0.2$) $\times10^{-13}$ erg s$^{-1}$ cm$^{-2}$. Given the weak ($< 2\sigma$) evidence for a thermal component and its minimal effect on the total flux, we prefer the simpler power-law model, but note that some small thermal contribution could be present.

The hard power law, with $\Gamma=1.74\pm0.04$, matches the X-ray spectrum of PSR J1023+0038 \citep{Bogdanov15} in the sub-luminous disk state as well as that of the tMSP candidates J1544.6--1125 \citep{Bogdanov16}, 3FGL J0427.9--6704 \citep{Strader16,Li20}, and CXOU J110926.4--650224 \citep{CotiZelati19}. A similar hard power law has also been observed for quiescent accreting millisecond pulsars such as SAX J1808.4--3658 \citep{Heinke09}, though such sources do not show an optical disk nor the extreme variability observed for tMSPs in the sub-luminous disk state.

\subsubsection{Relative X-ray and $\gamma$-ray flux}
\label{sec:fxfb}

\begin{figure}[ht!]
	\includegraphics[width=1.05\linewidth]{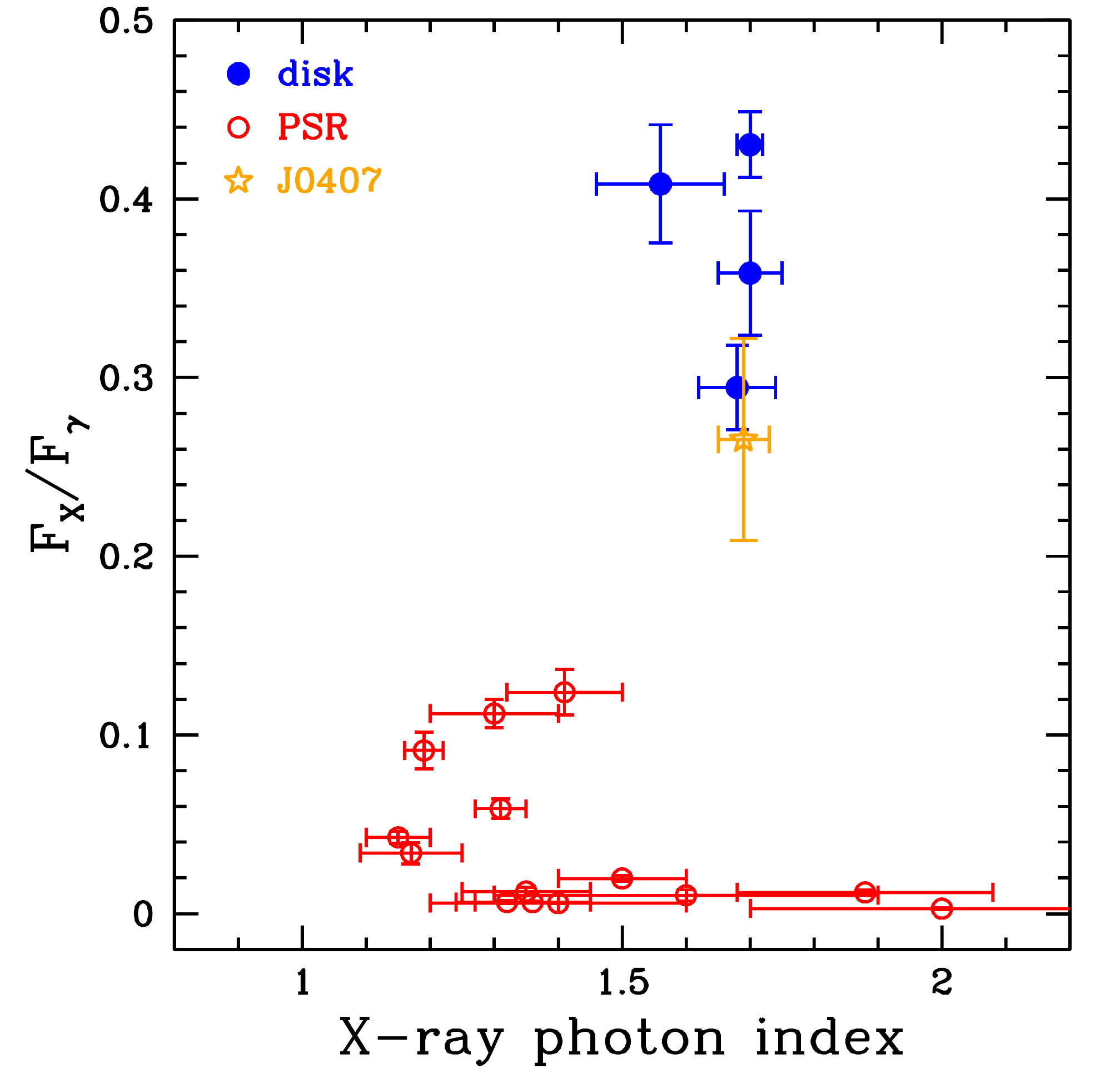}
    \caption{Ratio of X-ray (0.5--10 keV) to $\gamma$-ray (0.1--100 GeV) flux vs. X-ray photon index for tMSPs or candidates in the sub-luminous disk state (filled blue circles) and redbacks or tMSPs in the pulsar state (open red circles). The location of 4FGL J0407.7--5702 (orange star) is consistent with a classification as a disk state tMSP. The $\gamma$-ray fluxes are from \citet{4FGLDR2} and the X-ray fluxes from the compilation of \citet{Strader19}, except for PSR J1023+0038 \citep{Bogdanov11,Bogdanov15,Stappers14}, XSS J12270--4859 \citep{Linares_solo14,Johnson15}, and the new redback candidate 4FGL J2333.1--5527 \citep{Swihart20}. The X-ray photon indices are from the compilations of \citet{Linares_solo14} and \citet{Lee18} or the literature \citep{Bogdanov11,Bogdanov15,Bogdanov16,Li16,Strader16,Halpern17,AlNoori18,Cho18,Gentile18,Li18,Linares18,Swihart18,deMartino20,Li20,Swihart20}.}
    \label{fig:fxfg}
\end{figure}

Since the distance to 4FGL J0407.7--5702 is not  known, we cannot effectively assess whether its X-ray luminosity supports an identification as a candidate tMSP. Nevertheless, here we show that the ratio of its X-ray and $\gamma$-ray fluxes---a distance-independent quantity---does indeed support this classification.

The X-ray (0.5--10 keV) to $\gamma$-ray (0.1--100 GeV) flux ratio for 4FGL J0407.7--5702 is $F_X/F_{\gamma} = 0.26\pm0.06$. In Figure~\ref{fig:fxfg} we show this ratio, plotted against the X-ray photon index from a power-law spectral fit, for 4FGL-detected sources in the sub-luminous disk state\footnote{CXOU J110926.4--650224 was associated with a tentative 8-year source (FL8Y J1109.8--6500) that is not in the official 4FGL catalog. Using FL8Y, $F_X/F_{\gamma} = 0.67\pm0.27$.} (PSR J1023+03038, XSS J12270--4859, 3FGL J1544.6--1125, 3FGL J0427.9--6704) as well as all redbacks in the pulsar state that appear in the 4FGL catalog and which have well-measured photon indices (uncertainties $< 0.5$). 

Figure~\ref{fig:fxfg} shows that the four confirmed or candidate field tMSPs in the sub-luminous disk state have $F_X/F_{\gamma}$ in the range 0.29--0.43. This can be compared to a median value of 0.012 for redbacks in the pulsar state, and a maximum of $F_X/F_{\gamma} \sim 0.12$ (for 1FGL J1417.7--4407, which has an evolved companion and perhaps an unusually luminous intrabinary shock). 4FGL J0407.7--5702 has an $F_X/F_{\gamma}$ value consistent within the uncertainties with that observed for tMSPs, supporting its classification as such. By contrast, the recently discovered binary 4FGL J0935.3+0901, whose optical and X-ray data alone do not allow a clear classification \citep{Wang20}, has $F_X/F_{\gamma} \sim 0.02$, suggesting it is much more likely to be in the pulsar state than the sub-luminous disk state.

The transitions of PSR J1023+03038 and XSS J12270--4859 show the proximate reason for this difference between the disk and pulsar states: in the disk state their 0.1--100 GeV $\gamma$-ray fluxes are higher by only a factor of $\sim 3$--6 compared to the pulsar state \citep{Stappers14,Johnson15}, but their 0.5--10 keV X-ray fluxes are higher by a factor of $\sim 25$--30 \citep{Bogdanov11,Bogdanov15,Linares14,deMartino20}. Whether this difference holds for other tMSPs awaits future observed transitions, but the location of candidate disk-state tMSPs in the same region as confirmed tMSPs is suggestive.

If tMSPs do indeed all have relatively high values of $F_X/F_{\gamma}$, this has ramifications for the identification of new tMSPs among currently unassociated \emph{Fermi} $\gamma$-ray sources. New sources in the 10-year catalog have typical 0.1--100 GeV fluxes of 1--$3 \times 10^{-12}$ erg s$^{-1}$ cm$^{-2}$ \citep{4FGLDR2}. For typical tMSP-like values of $F_X/F_{\gamma}$, the corresponding 0.5--10 keV fluxes are $F_X \sim 3 \times 10^{-13}$ to $10^{-12}$ erg s$^{-1}$ cm$^{-2}$, detectable with \emph{Swift}/XRT for short exposure times of 1--2 ksec even in the presence of moderate foreground extinction ($N_H \lesssim 10^{22}$ cm$^{-2}$). The expected all-sky X-ray sensitivity of \emph{eROSITA} is similar \citep{Merloni12}. The implication is that for essentially any tMSP in the sub-luminous disk state detected as a \emph{Fermi} GeV source, an X-ray counterpart should be readily identifiable even in shallow data. This is unlike the case for redbacks or black widows, which can have much fainter X-ray counterparts. Since typical \emph{Fermi} error ellipses can contain a number of unrelated X-ray sources, the mere existence of a candidate X-ray counterpart cannot be used to provide a definitive classification, but the absence of such a counterpart would disfavor the identification of the source as a tMSP.

Figure~\ref{fig:fxfg} also shows that, consistent with previous work (e.g., \citealt{Linares_solo14}), the tMSPs or candidates in the disk state have $\Gamma \sim 1.7$, but that redbacks show a wide range of photon indices, in part depending on the strength of the intrabinary shock. Therefore, the X-ray photon index can only give indicative, but not conclusive, information about the classification of a candidate tMSP.

\subsubsection{ROSAT}

There is a faint ($0.020\pm0.008$ ct s$^{-1}$) \emph{ROSAT} source, 2RXS J040730.2--570024 \citep{Boller16}, with a catalog position $11\arcsec$ from that of the optical/X-ray counterpart to 4FGL J0407.7--5702. Assuming the best-fit \emph{XMM} spectral model, this \emph{ROSAT} count rate is equivalent to a 0.5--10 keV flux of $(3.6\pm1.4) \times10^{-13}$ erg s$^{-1}$ cm$^{-2}$, identical to the \emph{XMM} flux within the uncertainties. Given how similar these fluxes are, the lack of another \emph{XMM} X-ray source within $\sim 45\arcsec$ of the target, and the poor astrometry expected for faint \emph{ROSAT} sources, we think it is likely that 2RXS J040730.2--570024 is the same as the \emph{XMM} source despite the astrometric offset. If so, this implies that 4FGL J0407.7--5702 was in a similar spectral state in 1990--1991 to 2019--2020. While the $\gamma$-ray source is faint, there is no evidence for significant $\gamma$-ray variability since 2008 \citep{4FGLDR2}, and hence no reason to believe a transition occurred in this time interval. The constraints on a possible ``full" X-ray outburst are weak: at our inferred range of likely distances, an outburst to $L_X \sim 10^{36}$ erg s$^{-1}$ would have only reached a few milliCrab, so its discovery by all-sky X-ray monitors would have been borderline.

\subsection{X-ray Light Curve}
\label{sec:XrayLC}

\begin{figure}[t!]
\includegraphics[width=\linewidth]{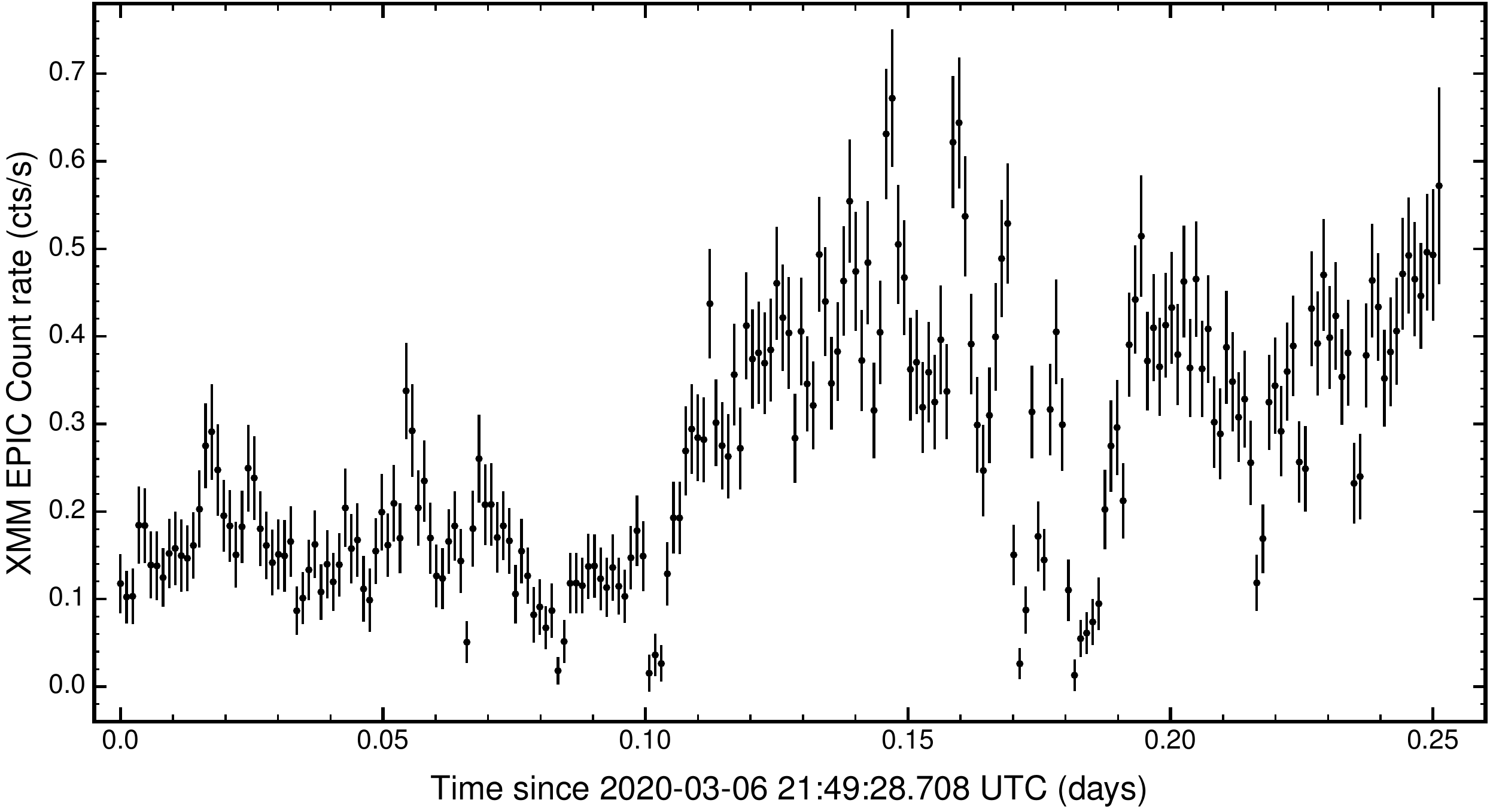}
\caption{Background subtracted and barycentric corrected \textit{XMM-Newton} EPIC light curve of 4FGL J0407.7--5702 in the 0.2--10 keV band, binned in 100s bins.}
\label{fig:xmm_bigLC}
\end{figure}

\begin{figure}[t]
\includegraphics[width=\linewidth]{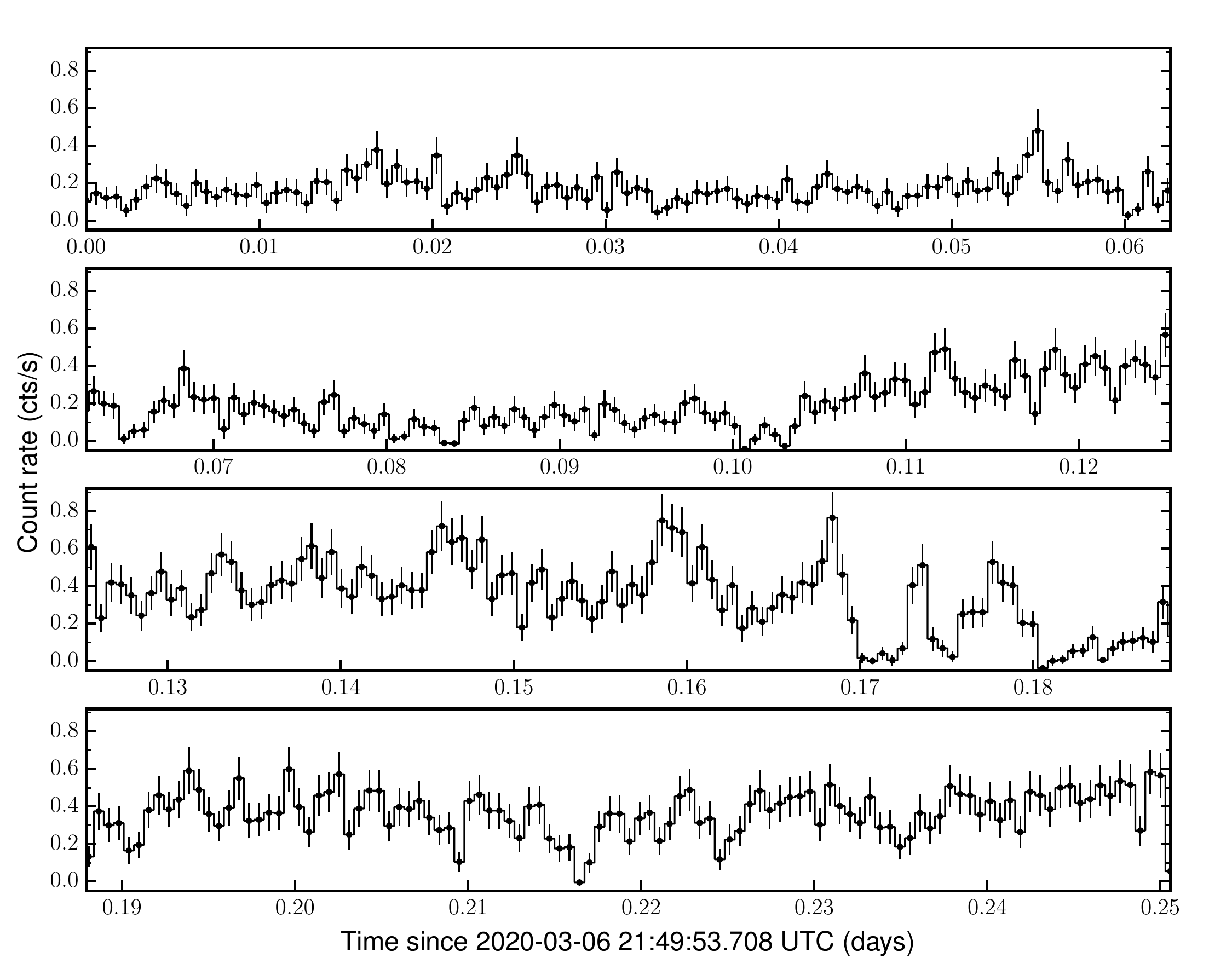}
\caption{The same data as in Figure 4, but instead with finer 50s bins.}
\label{fig:xmm_smallLC}
\end{figure}

\begin{figure}[t!]
\includegraphics[width=\linewidth]{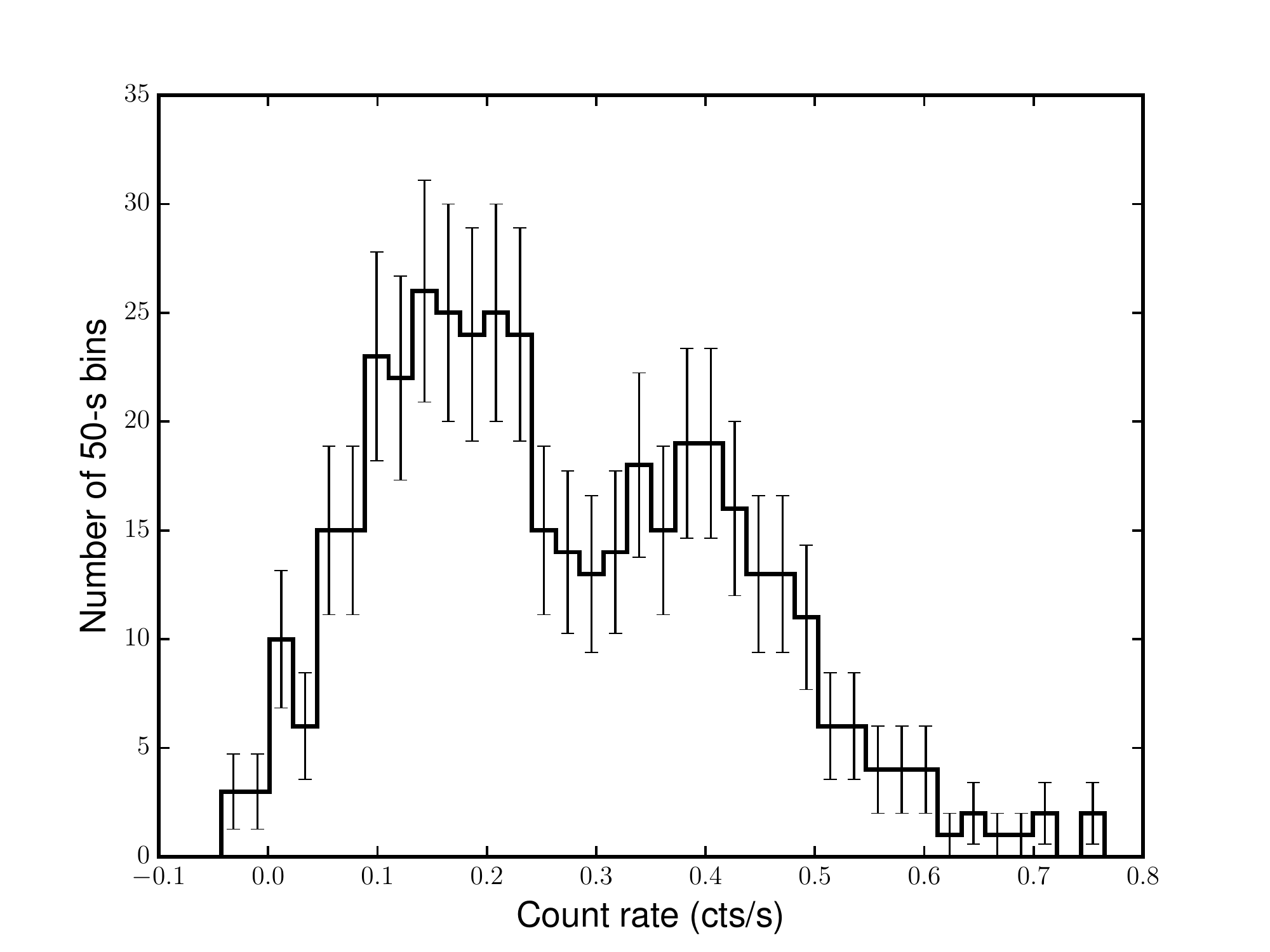}
\caption{Distribution of count rates from the 50s binned background subtracted \textit{XMM-Newton} EPIC light curve. We define the following regions according to their flux levels: low (0.0-0.3 ct s$^{-1}$), medium (0.3-0.6 ct s$^{-1}$), and flare ($>$0.6 ct s$^{-1}$).}
\label{fig:xmm_hist}
\end{figure}

The background subtracted and barycentric corrected \textit{XMM-Newton} EPIC 0.2--10 keV light curve, shown separately in 100s bins (Figure~\ref{fig:xmm_bigLC}) and 50s bins (Figure~\ref{fig:xmm_smallLC}), display clear short-term variability over the 21.7 ksec exposure. A histogram of the finer 50s-binned light curve shows a bimodal distribution (Figure~\ref{fig:xmm_hist}). For exploratory analysis in this section, we use this figure to guide a preliminary division of the light curve into three separate flux levels: low (0.0--0.3 ct s$^{-1}$), medium (0.3--0.6 ct s$^{-1}$), and flare ($>$0.6 ct s$^{-1}$). The average  0.2--10 keV count rate is $0.204\pm0.003$ ct s$^{-1}$.

4FGL J0407.7--5702 spent a majority of the observation at the low ($\sim$60\% of the time) or medium ($\sim$37\%) flux levels, with occasional flares ($\sim 2$--3\%). Given that fast, frequent ``mode switching" between the low and high modes is observed clearly in PSR J1023+0038, XSS J12270--4859, and 3FGL J1544.6--1125 \citep{Linares_solo14,Bogdanov15,BH15}, it is tempting to associate the low and medium count levels observed for 4FGL J0407.7--5702 with these well-studied modes.

However, there are several reasons to think this simple interpretation is not correct.
First, in these other systems the flux difference between the low and high modes is large (a factor of $\sim 5$--10), while in 4FGL J0407.7--5702 the difference is only a factor of $\sim 2$--2.5. Another difference is that in the other tMSPs and candidates the binary is in the high mode for the majority ($\gtrsim 75\%$) of the time, compared to $< 40\%$ here.

A careful examination of Figure~\ref{fig:xmm_smallLC} shows that the system does indeed make excursions to a count rate much fainter than the broad, low flux level identified in Figure~\ref{fig:xmm_hist}. The most extensive of these is around 0.17 d after the light curve start, where the binary has a flat-bottomed light curve with a mean count rate of $0.027\pm0.014$ ct s$^{-1}$. This is a factor of $\sim 7$--8 fainter than the average count rate over the whole dataset, and equivalent to an 
0.5--10 keV X-ray luminosity of $(3.2\pm1.6) \, (d/7 \, {\rm kpc})^2 \times 10^{32}$ erg s$^{-1}$. We suggest it might be more accurate to view this rarer state as the true ``low mode" and both the peaks at 0.15--0.2 ct s$^{-1}$ and 0.4 ct s$^{-1}$ as manifestations of the same ``high mode".

Indeed, from a phenomenological point of view the X-ray light curve most closely resembles that of the tMSP candidate CXOU J110926.4--650224, which shows occasional low modes but less pronounced bimodality over its entire light curve than some of the other tMSPs \citep{CotiZelati19}. It may also be the case that tMSPs show a broader set of behaviors than simple mode switching; for example, the tMSP candidate 3FGL J0427.9--6704 shows bright $\gamma$-ray and radio continuum emission as well as a disk, appears to spend most of its time in an X-ray flare mode, with no consistent stable low or high modes \citep{Li20}.

\subsection{Optical Spectroscopy}
\label{sec:optspec}

\begin{figure*}
	\includegraphics[width=1.0\linewidth]{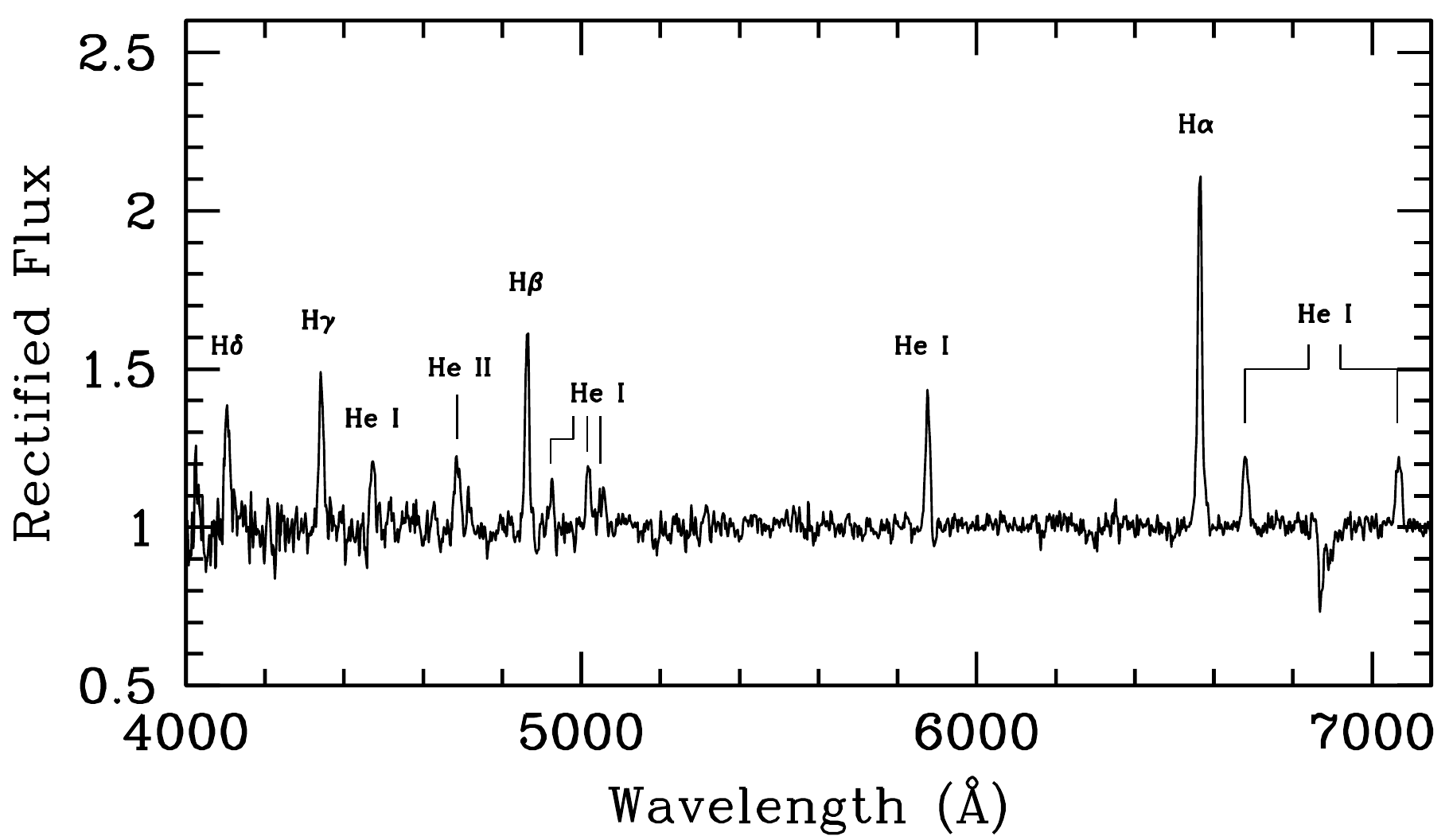}
    \caption{Rectified co-added SOAR optical spectra of 4FGL J0407.7--5702. The main Balmer and \ion{He}{1} and \ion{He}{2} lines are marked. The spectrum is that of a typical accretion disk, with the only significant absorption lines telluric.}
    \label{fig:opt_spectra}
\end{figure*}

All of the SOAR spectra show the same features: a blue continuum with superimposed emission lines from \ion{H}{1}, \ion{He}{1}, and the 4686 \AA\ line of \ion{He}{2}. The emission lines are resolved, with a mean resolution-corrected FWHM of $516\pm21$ km s$^{-1}$ for H$\alpha$ measured among the different epochs. A mean of the stronger \ion{He}{1} lines is yet broader at $\sim 670$ km s$^{-1}$, and the \ion{He}{2} 4686 \AA\ line is double-peaked with a FWHM $\sim 830$ km s$^{-1}$. This trend of increasing 
FWHM is consistent with the idea that the H emission is primarily from the outer disk, with \ion{He}{1} and then \ion{He}{2} dominated by regions progressively closer to the compact object.

There are no photospheric absorption features apparent either visually or in a cross-correlation with templates of the expected spectral types of the likely low-mass secondary. We obtained the Gemini spectra in the hope of uncovering faint absorption features in higher signal-to-noise spectra, but did not find any in these data either.

We next attempted to constrain the orbital period through motion of the emission lines. While the different epochs of data do show evidence for modest variations in the wavelengths of the emission lines (of order $\sim 20$--30 km s$^{-1}$), these did not phase on any readily identifiable orbital period.

Qualitatively, in terms of the presence of emission lines and their relative strengths, the 4FGL J0407.7--5702 spectra are very similar to those of the confirmed tMSP PSR J1023+0038 in its disk state and 
to the candidate tMSP 3FGL J1544.6--1125, and consistent overall with the optical spectra expected for an accretion disk around a compact object.

Under the assumption that the emission lines do arise in an accretion disk around a neutron star, their relatively narrow FWHM hints at a more face-on inclination. For example, the neutron star low-mass X-ray binary Cen X-4 has an H$\alpha$ FWHM of $678\pm48$ km s$^{-1}$ \citep{Casares15} and an inclination of $35_{-4}^{+5^{\circ}}$ \citep{Hammerstein18}, while 3FGL J1544.6--1125 has a FWHM of $\sim 330$ km s$^{-1}$ and an inclination of 5--$8^{\circ}$ \citep{Britt17}. 4FGL J0407.7--5702, with an H$\alpha$ FWHM of $516\pm21$ km s$^{-1}$, likely has an inclination within this broad range of face-on values, assuming its orbital period and primary mass are not too dissimilar from typical neutron star low-mass X-ray binaries. 

\subsection{Optical Photometry}
\begin{figure}[htbp]
\centering
\subfloat{\includegraphics[angle=0,width=0.47\textwidth,
trim={0 0.0cm 0 0},clip]{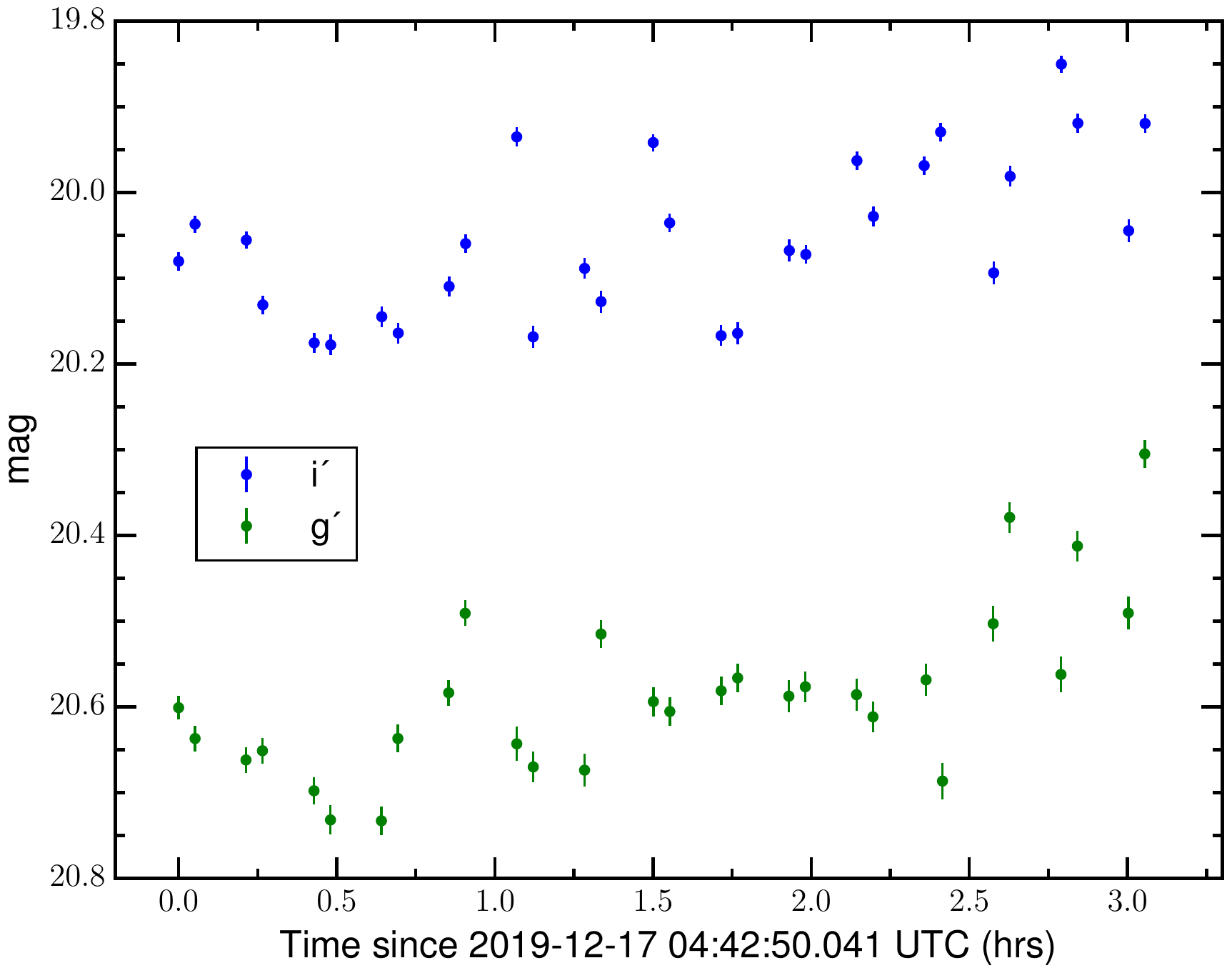}}\\
\subfloat{\includegraphics[angle=0,width=0.47\textwidth,
trim={0 0.0cm 0 0},clip]{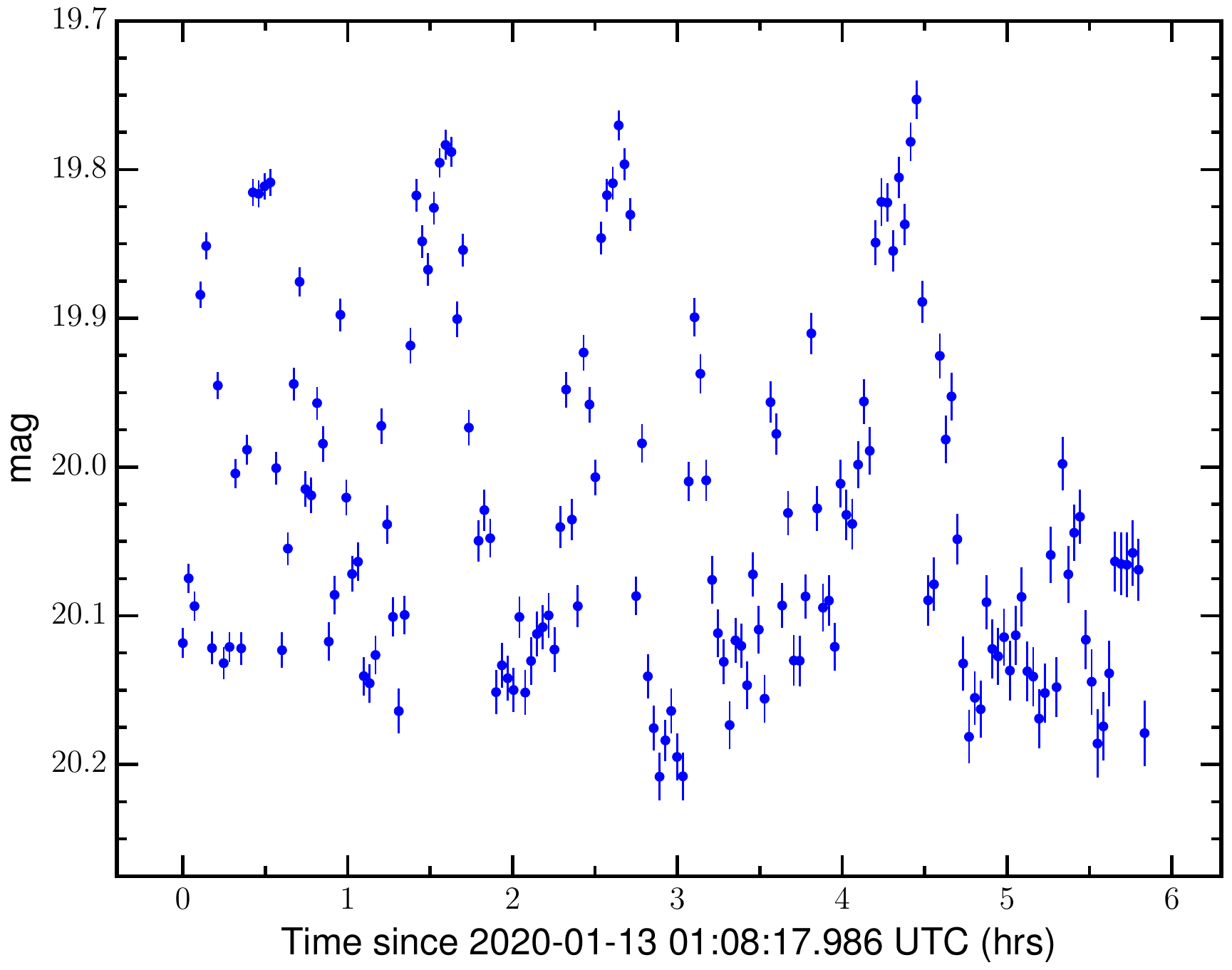}}
\caption{Top: SOAR $g'$ and $i'$ optical photometry of 4FGL J0407.7--5702, taken on 2019 Dec 17. Bottom: SOAR $i'$ photometry from 2020 Jan 13, showing the limit cycle behavior discussed in the text.}
\label{fig:opt_LCs}
\end{figure}

The SOAR optical ($g'$ and $i'$) light curves from our two epochs are shown in Figure~\ref{fig:opt_LCs}. In the
Dec 2019 epoch, with $g'$ and $i'$ data taken over three hours, the light curves in both filters show aperiodic, seemingly stochastic variations. The short timescale variations in both filters are reminiscent of the ``flickering'' seen in the light curve of PSR J1023+0338 while in its sub-luminous disk state \citep{Kennedy18}.

The Jan 2020 epoch, with data only in $i'$, extends over a longer timespan of nearly 6 hr. The variability is qualitatively different than the earlier optical photometry. By eye there is some evidence for periodic variability in the repeating distinct maxima (Figure~\ref{fig:opt_LCs}), but these do not repeat regularly: the second and third peaks are separated by $\sim 1.1$ hr, and the third and fourth by $\sim 1.7$ hr. A Lomb-Scargle periodogram analysis of this light curve does not show evidence for a significant peak (with only a weak, insignificant peak at 56 min). It is unlikely that any of these timescales represent the orbital period of the binary; of the known redbacks, the system with the shortest well-measured
orbital period is PSR J1622--0315, at 3.9 hrs \citep{Sanpa16}. It is unknown whether all the tMSP candidates are indeed redbacks and hence whether less massive donors and hence shorter periods might be possible.

Instead, the 2020 Jan 13 light curve more closely resembles the ``limit cycle'' behavior present most clearly in the tMSP candidate 3FGL J1544.6--1125, which shows short timescale variations confined between minimum and maximum values separated by $\sim$0.5 mag \citep{BH15}, as opposed to random flickering. Similar behavior is also observed in XSS J12270--4859 \citep{Pretorius2009,deMartino10} and PSR J1023+0038 \citep{Bond02,Bogdanov15,Kennedy18}, although in these systems modulation on the orbital period is also observed, superimposed on the shorter timescale variations. Such orbital variations, if present, might be harder to observe for 4FGL J0407.7--5702 given its likely face-on orientation (Sec.~\ref{sec:optspec}).

\subsection{Distance}
\label{sec:distance}

Since 4FGL J0407.7--5702 does not have a significant \emph{Gaia} DR2 parallax ($\varpi = -0.37\pm0.48$ mas, \citealt{GaiaDR2}), nor the possibility  of modeling the flux from a Roche Lobe-filling secondary, we must use alternative methods to constrain its distance. To estimate the most likely distance, we proceed under the assumption that it has intrinsic properties similar to the known tMSPs. 

First, we consider its $\gamma$-ray and X-ray flux in the context of the four known or candidate tMSPs that have reasonably well-constrained distances. From the compilation of \citet{Strader19}, the 0.1--100 GeV $\gamma$-ray luminosities of these four sources range from  $6 \times 10^{33}$ erg s$^{-1}$ to $2.4 \times 10^{34}$ erg s$^{-1}$, which given the flux of 4FGL J0407.7--5702 would imply a distance in the range 5.6--11.2 kpc. Similarly, given mean 0.5--10 keV X-ray luminosities of $2.4 \times 10^{33}$ erg s$^{-1}$ to $7.7 \times 10^{33}$ erg s$^{-1}$ for these sources, the implied X-ray distance is in the range 6.9--12.5 kpc.

As a third  estimate we consider the \emph{Gaia} photometry for the three sources with well-constrained distances and which have been in the disk state for the \emph{Gaia} era: PSR J1023+0038, 3FGL J1544.6--1125, and 3FGL J0427.9--6704. We focus on the $G_{BP}$ photometry under the hypothesis that this region of the spectrum is more likely to be disk-dominated, and under the ansatz that given the vaguely similar X-ray luminosities and periods of these sources their blue disk luminosities might also
be similar. Using \emph{Gaia} DR2 \citep{GaiaDR2}, accounting for foreground reddening \citep{Schlafly11,Marigo17}, and the distances from \citet{Strader19}, the range of absolute $M_{G_{BP}}$ for this small sample of objects is indeed small, from $M_{G_{BP}} \sim 5.6$ to 6.0. Given $G_{BP,0} = 20.22 \pm 0.09$ for 4FGL J0407.7--5702, the implied range of optical distances is 7.2--8.3 kpc. The systematic uncertainties associated with this optical distance estimate are likely much larger than even those for the X-ray and $\gamma$-ray distances, but nevertheless, these values fall within the the range of the high-energy distances.

Above and for the remainder of the paper we quote a reference distance of 7 kpc, but emphasize that this is not a best-fit distance estimate of the binary and that the uncertainty on the distance is substantial. We are more secure in saying that if 4FGL J0407.7--5702 has intrinsic properties similar to the other known and candidate tMSPs, then a distance of  $\lesssim 5$ kpc is disfavored.

\section{Discussion and Conclusions}
\label{sec:discussion}
We have shown that the brightest X-ray source within the error ellipse of the \emph{Fermi}-LAT $\gamma$-ray source 4FGL J0407.7--5702 has an X-ray light curve and spectrum consistent with known and strong candidate tMSPs in the sub-luminous disk state. Photometry and spectroscopy of the optical counterpart to this X-ray source provides compelling support for this scenario, as does the high ratio of X-ray to $\gamma$-ray flux.

A definitive classification as a tMSP would require an observed transition to the pulsar state, which could be clued by $\gamma$-ray, X-ray, and optical monitoring of the source. Nonetheless, our conclusion could be strengthened with additional data in the present state.
The most straightforward of these would be longer X-ray observations with \emph{XMM-Newton} or \emph{Chandra}, which could reveal whether the hints of mode switching
behavior discussed above are borne out with more data. X-ray timing observations with one of these telescopes, or perhaps with \emph{NICER}, could in principle reveal if the system shows X-ray pulsations similar to PSR J1023+0038 in the sub-luminous disk state \citep{Jaodand16}, though in the absence of radio ephemerides such detections are challenging. Alternatively, it is possible that 4FGL J0407.7--5702 shows X-ray phenomenology beyond mode switching, not unlike 3FGL J0427.9--6704, supporting a wider range of behaviors for these systems. 

Another useful measurement would be the orbital period of the binary, which we could not determine using our optical photometry or spectroscopy. Rapid cadence optical photometry, or photometry or spectroscopy in the near-infrared (where the contribution of the donor star might be more observable compared to the disk) are potential alternative approaches.

Since PSR J1023+0038 \citep{Deller15}, XSS J12270--4859 \citep{Hill11}, 3FGL J1544.6--1125 \citep{Jaodand19}, and 3FGL J0427.9--6704 \citep{Li20} all show radio continuum emission in their sub-luminous disk states, the radio behavior of 4FGL J0407.7--5702 could strengthen its tentative classification as a candidate tMSP. The plausibility of a detection depends on its unknown distance and radio behavior: if at 7 kpc, it would be well-detectable if at the 5 GHz radio luminosity of 3FGL J0427.9--6704 (predicted flux density of $\sim 31 \mu$Jy), marginal if as luminous as XSS J12270--4859 or 3FGL J1544.6--1125, and very difficult if akin to PSR J1023+0038 (mean flux density $\sim 2$--6 $\mu$Jy, though with occasional brighter flares).

Under the assumption that 4FGL J0407.7--5702 has a luminosity comparable to previously studied tMSPs in the X-ray, $\gamma$-ray, or optical, we found the most likely distance to lie in the range $\sim 6$--13 kpc, with a consensus value more toward the lower end of that range. While the uncertainty in these estimates is substantial, together they suggest that 4FGL J0407.7--5702 is the most distant known candidate field tMSP to date. It is possible, though not certain, that an end-of-mission \emph{Gaia} parallax for this source might be available, which would allow the crucial determination of its X-ray luminosity.

Finally, we highlight the emerging evidence that the ratio of X-ray to $\gamma$-ray flux ($F_X/F_\gamma$) could help to identify candidate tMSPs in the sub-luminous disk state and separate them from the more common redbacks or black widows when good distance constraints are not available. Since the increasingly deep \emph{Fermi}-LAT catalogs are enabling the study of more distant millisecond pulsar binaries, such techniques may see increasing relevance in follow-up studies in the coming years.

\section*{Acknowledgements}
We acknowledge a helpful and thoughtful report from an anonymous referee. We also acknowledge support from NSF grant AST-1714825 and the Packard Foundation.

This work is based on observations obtained with \textit{XMM-Newton} an ESA science mission with instruments and contributions directly funded by ESA Member States and NASA. 

Based on observations obtained at the Southern Astrophysical Research (SOAR) telescope, which is a joint project of the Minist\'{e}rio da Ci\^{e}ncia, Tecnologia, Inova\c{c}\~{o}es e Comunica\c{c}\~{o}es (MCTIC) do Brasil, the U.S. National Optical Astronomy Observatory (NOAO), the University of North Carolina at Chapel Hill (UNC), and Michigan State University (MSU).

Based on observations obtained at the international Gemini Observatory, a program of NSF’s OIR Lab, which is managed by the Association of Universities for Research in Astronomy (AURA) under a cooperative agreement with the National Science Foundation. on behalf of the Gemini Observatory partnership: the National Science Foundation (United States), National Research Council (Canada), Agencia Nacional de Investigaci\'{o}n y Desarrollo (Chile), Ministerio de Ciencia, Tecnolog\'{i}a e Innovaci\'{o}n (Argentina), Minist\'{e}rio da Ci\^{e}ncia, Tecnologia, Inova\c{c}\~{o}es e Comunica\c{c}\~{o}es (Brazil), and Korea Astronomy and Space Science Institute (Republic of Korea).

\software {IRAF \citep{Tody86, 2016ascl.soft08006G}, SAS \citep[v18.0.0,][]{Gabriel04}, Astropy \citep{astropy13}}

\bibliography{main}

\end{document}